\documentclass{aa}
\usepackage{psfig}

\def\bron{SAX~J1748.9-2021}
\def\ecs{erg~cm$^{-2}$s$^{-1}$}
\def\lum{erg~s$^{-1}$}

\begin{document}
\thesaurus{05(08.09.2 \bron; 08.09.2 MX~1746-20; 08.14.1; 10.07.3 NGC~6440; 13.25.1; 13.25.5)}

\title{A new X-ray outburst in the globular cluster NGC~6440: \bron}
\author{J.J.M.~in~'t~Zand\inst{1} 
 \and F.~Verbunt\inst{2}
 \and T.E.~Strohmayer\inst{3}
 \and A.~Bazzano\inst{4}
 \and M.~Cocchi\inst{4} 
 \and J.~Heise\inst{1}
 \and M.H.~van~Kerkwijk\inst{2}
 \and J.M.~Muller\inst{1,5} 
 \and L.~Natalucci\inst{4}
 \and M.J.S.~Smith\inst{1,6}
 \and P.~Ubertini\inst{4}}
\offprints{J.J.M.~in~'t Zand (at e-mail {\tt jeanz@sron.nl})}

\institute{     Space Research Organization Netherlands, Sorbonnelaan 2,
                NL - 3584 TA Utrecht, the Netherlands
         \and
                Astronomical Institute,
                P.O.Box 80000, NL - 3508 TA Utrecht, the Netherlands
	 \and
	 	NASA Goddard Space Flight Center, Code 666, Greenbelt, MD 20771, 
		U.S.A.
         \and
                Istituto di Astrofisica Spaziale (CNR), Area Ricerca Roma Tor
                Vergata, Via del Fosso del Cavaliere, I - 00133 Roma, Italy
         \and
                {\em BeppoSAX\/} Science Data Center, Nuova Telespazio,
                Via Corcolle 19, I - 00131 Roma, Italy
         \and
                {\em BeppoSAX\/} Science Operation Center, Nuova Telespazio,
                Via Corcolle 19, I - 00131 Roma, Italy
                        }
\date{Received 11 December 1998, accepted 10 February 1999}
\maketitle

\begin{abstract}
For the second time in 27 years a bright transient X-ray source has been 
detected coincident with the globular cluster NGC~6440. It was found to be
active in August, 1998, with the Wide Field Camera and the narrow field 
instruments on the {\em BeppoSAX\/} spacecraft, and with the All-Sky 
Monitor and
the Proportional Counter Array on the {\em RossiXTE\/} spacecraft. Four X-ray 
bursts were detected, at least one of which shows the characteristics of a
thermonuclear flash on a neutron star, in analogy with some $\sim20$
optically identified low-mass X-ray binaries.
The broad-band spectrum is hard as is common among low-mass X-ray binaries of 
lower luminosity ($\la10^{37}$~erg~s$^{-1}$) and can be explained by a 
Comptonized model. During the burst the $>30$~keV emission brightened,
consistent with part of the burst emission being Compton up scattered
within $\sim10^{11}$~cm.
\end{abstract}

\section{Introduction}
\label{intro}

If the ratio of the number of bright ($\ga10^{-9}$~\ecs\ in 2 to 10 keV) X-ray 
sources to that of
ordinary stars in Galactic globular clusters were the same as in the Galactic
disk one would expect to find 0.1 bright X-ray source in all $\sim150$ clusters
combined. In reality, 12 are known. Comparison between these 12 sources
and the bright X-ray sources in the Galactic disk sets the two populations
further apart: for instance, $\sim85$\% of the globular cluster sources 
are X-ray bursters against only $\sim20$\% in the disk, none of the transient
X-ray sources in globular clusters is a strong black hole candidate, many are in
the disk. These facts point to different evolutionary scenarios between both
populations. The different
average ages of both populations can only partly explain the different
characteristics.

A few of the 12 bright X-ray sources in globular clusters are transient, implying 
they are only occasionally bright. One of these was detected in December 1971 in
NGC~6440 (Markert et al. 1975). NGC~6440 is located close to the Galactic center,
at a distance of $8.5\pm0.4$~kpc from us and 0.6~kpc above the Galactic plane 
(Martins et~al. 1980,
Ortolani et~al. 1994). The distance is based on V and I photometry of 
horizontal branch stars in the color-magnitude diagram (Ortolani et al. 1994).
The distance scale to globular clusters is currently being revised (e.g.,
Chaboyer et al. 1998); we will adhere to the distance of 8.5~kpc
in this paper. NGC~6440 is fairly dense and apparently its core has not 
undergone a collapse, the core radius is 7.6\arcsec\ and the tidal radius 
6.3\arcmin\ (Trager et~al. 1993). Apart from the 1971 transient, 
one faint X-ray source
was identified with ROSAT in NGC~6440 by Johnston et~al. (1995).
This source may or may not be related to the transient.

\begin{figure}[t]
\psfig{figure=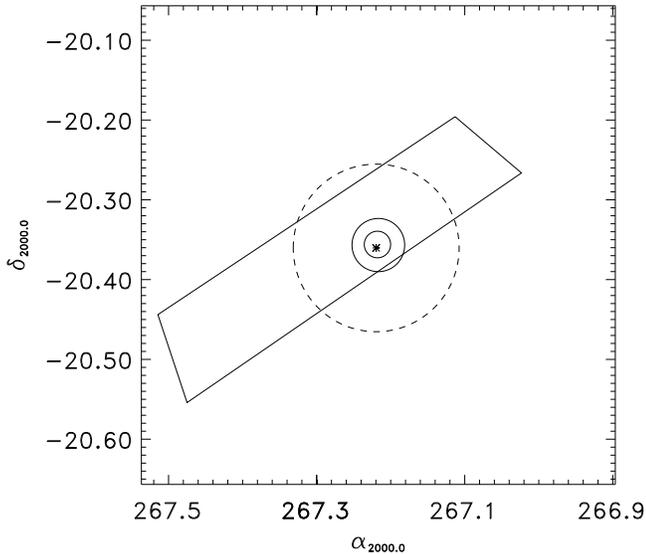,width=\columnwidth,clip=t}

\caption[]{The large solid circle is the 99\% confidence error box from
the WFC observation (In 't Zand et al. 1998a), the small solid circle that
from the MECS observation. The quadrilateral is the combined Uhuru-OSO-7
error box for MX~1746-20 (Forman et~al. 1976). The asterisk
indicates the center of NGC~6440 (Picard \& Johnston 1995), it has a core
radius of 0\fdg002, the large dashed circle refers to the tidal radius. 
The coordinates are in degrees for equinox 2000.0

\label{figerr}}
\end{figure}

In this paper a new X-ray outburst in NGC~6440 is discussed.
It was discovered with the Wide Field Cameras (WFC) on the {\em BeppoSAX\/} 
spacecraft in August 1998 (In~'t~Zand et al. 1998a), almost
27 years after the previous one was noticed. The All-Sky Monitor
(ASM) on the {\em RossiXTE\/} spacecraft also made a weak detection.
Follow-up observations were performed with the 
narrow-field instruments (NFI) on {\em BeppoSAX\/} and with the Proportional
Counter Array (PCA) on {\em RossiXTE}.
We present the analysis of the observations with these four sets
of instruments and speculate on the nature of the source. 
The analysis of each data set concentrates on its specific qualities.
In Sect.~\ref{secobs}, we outline the discovery observations with WFC 
and study the long term behavior with WFC and ASM. In Sect.~\ref{secspec}, 
we discuss the broad-band spectrum of the persistent emission as well as a 
burst detected with the NFI. The PCA observation is discussed 
in Sect.~\ref{secvar}, it focuses on the short-term variability of the
source. We review the results and their implications in Sect.~\ref{secdis}.

\section{Discovery, bursts and long-term trend}
\label{secobs}

\begin{figure}[t]
\psfig{figure=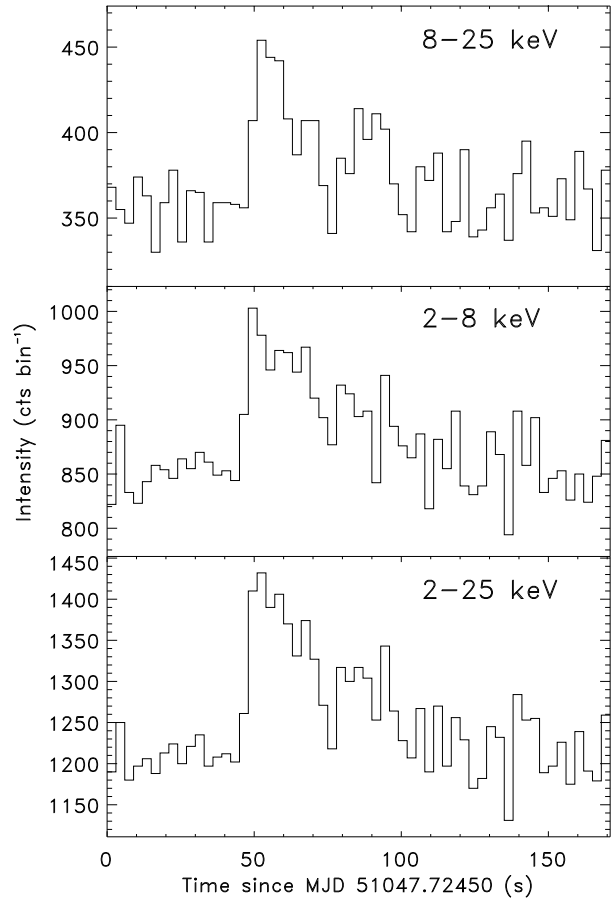,width=\columnwidth,clip=t}

\caption[]{Time profile of the second burst detected with WFC from \bron.
Photons were accumulated over only that part of the detector illuminated
by the source. No background was subtracted. The time resolution is 3~s
\label{figwfcburst}}
\end{figure}

\begin{table}[b]
\caption[]{Positions of NGC 6440 and coincident X-ray sources (Equinox 2000.0). 
$\epsilon$ is the uncertainty of the position quoted (better than 95\% 
confidence)}
\begin{tabular}{llll}
\hline
Source$^\ast$            & R.A.             & Decl.  & $\epsilon$ \\
\hline
NGC 6440 center$^{\rm a}$& 17h 48m 52.7s    & -20$^{\rm o}$ 21\arcmin 36.9\arcsec & \\
MX 1746-200$^{\rm b}$    & 267\fdg11        & -20\fdg20                         &  \\
\hspace{2mm}quadrilateral& 267\fdg02        & -20\fdg27                         &  \\
\hspace{2mm}coordinates  & 267\fdg47        & -20\fdg55                         &  \\
                         & 267\fdg51        & -20\fdg44                         &  \\
{\small \bron}$^{\rm c}$ & 17h 48m 52s      & -20$^{\rm o}$ 21\farcm4           & 2\arcmin \\
{\small \bron}$^{\rm d}$ & 17h 48m 53.4s    & -20$^{\rm o}$ 21\arcmin 43\arcsec & 1\arcmin \\
ROSAT$^{\rm e}$          & 17h 48m 51.4s    & -20$^{\rm o}$ 21\arcmin 51\arcsec & 5\arcsec \\
\hline
\end{tabular}
$^\ast$References: $^{\rm a}$Picard \& Johnston 1995; $^{\rm b}$Forman et~al. 1976;
$^{\rm c}$In 't Zand et al. 1998a; $^{\rm d}$this publication
$^{\rm e}$Johnston et al. 1995
\label{tabpositions}
\end{table}

The WFC (Jager et al. 1997) on the {\em BeppoSAX\/}
satellite (Boella et al. 1997a) is carrying out a program of monitoring
observations of the 40$^{\rm o}\times40^{\rm o}$ field around the Galactic
center. The purpose is to detect X-ray 
transient activity, particularly from low-mass X-ray binaries (LMXBs) whose
Galactic population exhibits a strong concentration in this field, and
to monitor the behavior of persistently bright X-ray sources. Some previous
results have been published in In~'t~Zand et al. (1998b), Heise
(1998), Cocchi et al. (1998).

The program consists of campaigns during the spring and fall of each year. 
Each campaign lasts about two months and typically comprises weekly
observations. Up to the spring 1998 campaign, a total net exposure
time of 2 million seconds was accumulated during 4 campaigns. The fall 1998
campaign consists of five observations of $\sim50$~ks each between August 
22 and October 7, 1998.

During the first observation, on August 22.40-23.14, 1998, a transient was 
detected at a position consistent with that of the globular cluster NGC~6440
with a flux of $\sim30$~mCrab (2-10~keV). The
centroid position was 0\farcm3 from the center of NGC~6440 (In~'t~Zand
et al. 1998a, Picard \& Johnston 1995). The source was designated \bron.
The error box is indicated in Fig.~\ref{figerr} and specified 
in Table~\ref{tabpositions}. The figure also includes
the error box of the previous transient coincident with NGC~6440: MX~1746-20
which was detected with instruments on board the Uhuru and OSO-7 spacecrafts
in December 1971 and January 1972 (Forman et~al. 1976). ROSAT and 
optical observations of NGC~6440 were made soon after the WFC detection
and will be dealt with in a separate paper (Verbunt et al., in preparation).

The source was detected with WFC only during the August 22, 1998, observation. 
The last WFC observation prior 
to this was carried out on April 7, 1998 (137 days before the detection) with 
a 3$\sigma$ upper limit of 5 mCrab (2-10 keV) at the position of \bron. The 
WFC observation following that of August 22 was performed on September 1, with
an identical upper limit. These upper limits are about 6 times lower than the 
level during the detection. Given the rather sporadic
coverage, we are unable to assess the duration of the transient activity
from WFC data alone. Therefore, we studied data from the all-sky monitor 
(ASM) on {\em RossiXTE\/} (Levine et al. 1996)\footnote{the ASM measurements are 
publicly available at URL 
{\tt http://heasarc.gsfc.nasa.gov/docs/xte/asm\_products.html}} which covers
the period MJD~50094 through 51177 (Jan. 12, 1996, through Dec. 30, 1998). 
The ASM is less sensitive than the WFC per day of observation but this is 
compensated by a rather 
uniform and continuous coverage in time and sky position. At the position of 
NGC~6440, the $3\sigma$ detection limit in 1 day of exposure is about 
30~mCrab in 2 to 10 keV but one should bear
in mind that this can vary by roughly a factor 2 due to observational
constraints. 

In order to search for a significant ASM signal from a source in
NGC~6440, we rebinned the data to a time resolution of 7~days. This
increases the sensitivity to about 10 mCrab while preserving at least some
information on the variability of the source and preventing that the 
transient signal smears out below the detection threshold.
We varied the phasing of the binning in steps of 1~day to try to optimize
the synchronization with the onset of the transient activity. A bias level
was subtracted in the light curve to account for contributions from other
non-transient sources (the region is likely to be source confused from time 
to time; GX~9+1 is only 3\degr\ from NGC~6440). 
This level was determined by weight averaging the data for periods at least
100~days from MJD~51047 (when it was detected with WFC). The result of
this data reduction is that a source at NGC~6440 was detected above the 
$\sim10$~mCrab detection threshold in one 7-day bin. The average intensity
in this bin is $51\pm8$~mCrab. During almost two years of coverage,
NGC~6440 was brightest in this particular bin. The bin starts at MJD~51044. 
In the next 7-day bin the source is below a $3\sigma$ upper limit of 
7~mCrab. Taking into account the last detection with NFI and PCA 
(Sects.~\ref{secspec} and \ref{secvar}, on MJD~51051) and the non-detection
with WFC on MJD~51057, we conclude that \bron\ was active at a level above 
5 mCrab between 7 and 12 days.

Three bursts were detected with WFC, on MJD~51047.60976, 51047.72506
and 51047.84226. It is intriguing that the wait times differ by only 
164~s or 1.7\%. This reminds one of the narrow distribution of wait 
times found in GS1826-238 by Ubertini et al. (1999).
A constant time interval between bursts in the case of \bron\
would lead to three other bursts during the
WFC observation, all of which coincide with Earth occultations
or with standby mode of the instruments during passage of {\em BeppoSAX\/}
through the South Atlantic Anomaly. Thus, we can neither confirm
nor exclude a strict periodicity of burst times during our WFC observation.
The bursts had peak intensities between 0.5 and 0.7 Crab units (2-10 keV).
This is relatively
faint for WFC and the time profiles (see Fig.~\ref{figwfcburst}) are
rather noisy so that it is difficult to detect a softening of the spectrum
during these bursts. The durations of the three bursts were similar: about
100~s in the 2 to 10 keV band. The average spectrum of the three bursts,
accumulated over 62~s of each burst (this covers both time intervals
employed in Sect.~\ref{secburst}), is consistent with a black body
radiation model with a temperature of $kT=2.31\pm0.18$~keV 
and a radius of $(4.9\pm0.8)d_{\rm 8.5~kpc}$~km ($\chi^2_{\rm r}=0.8$ for 69 dof;
$d_{\rm 8.5~kpc}$ is the distance in units of 8.5~kpc).
Since a similar burst was detected with the more sensitive NFI 
as well (see Sect.~\ref{secspec}), it was decided not to analyze the 
WFC-detected bursts any further.

The average spectrum of \bron\ as measured with WFC is consistent with a 
power-law model with a photon index of $1.8\pm0.2$ and absorption 
by interstellar gas with cosmic abundances with a hydrogen column density 
of $N_{\rm H}=(4.0\pm1.6)\times10^{22}$~cm$^{-2}$ (the cross 
section per hydrogen atom was taken from Morrison \& McCammon 1983), 
the reduced $\chi^2$ is
$\chi^2_{\rm r}=0.78$ for 21 dof (whenever an error for a spectral parameter
is quoted throughout this paper, it refers to the single parameter 1$\sigma$
error). The average flux was 
$(7.0\pm0.4)\times10^{-10}$~\ecs\ in 2 to 10 keV. Furthermore, the shape of
the WFC-measured spectrum is also reasonably consistent with the Comptonized 
spectrum discussed in Sect.~\ref{secsppe} (for the spectral parameters, see 
Table~\ref{tabpersistentfit}): $\chi^2_{\rm r}=1.38$ for 23 dof.
No variability was detected
on time scales above 1~min, the 3$\sigma$ upper limit on a time scale of 5~h 
was $\sim$30\%.

\begin{figure}[t]
\psfig{figure=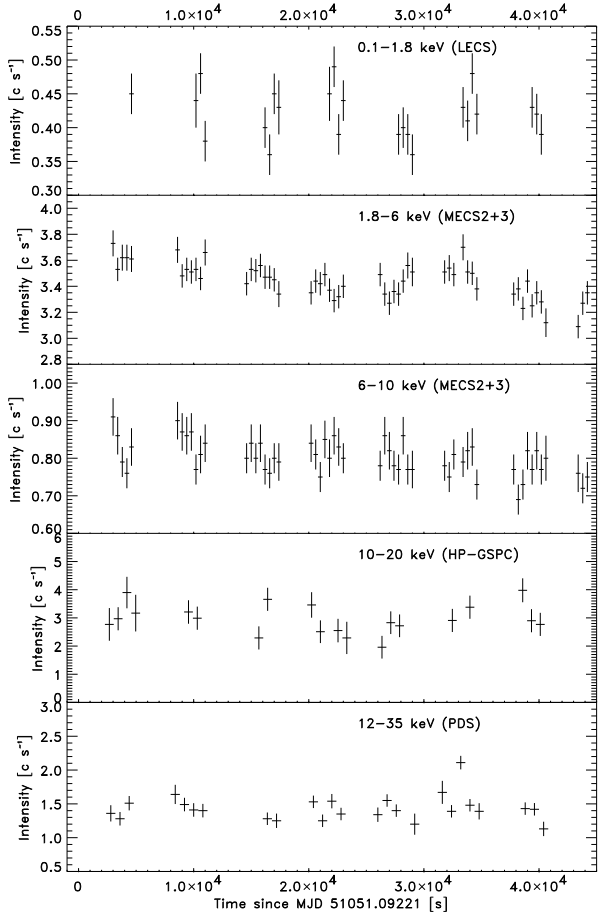,width=\columnwidth,clip=t}

\caption[]{Lightcurve as measured with the NFI in a number of bandpasses, 
corrected for background. The time resolutions are 400~s for the
upper three panels and 800~s for the lower two panels. Data for the burst were
excluded in these graphs. The background levels are from top to bottom
$6.1\times10^{-3}$, $2.6\times10^{-3}$, $1.9\times10^{-3}$, 30 and
3~c~s$^{-1}$ 
\label{figpersistentlc}}
\end{figure}

\section{Broad-band spectral measurements with {\em BeppoSAX\/}-NFI}
\label{secspec}

The NFI include 2 imaging instruments 
that are sensitive at photon energies below 10~keV: the Low-Energy and the 
Medium-Energy Concentrator Spectrometer (LECS and MECS, see 
Parmar et al. 1997 and Boella et al. 1997b respectively). They both have 
circular fields of view with diameters of 37 and 56 arcmin and effective 
bandpasses of 0.1-10 and 1.8-10~keV, respectively. The other two, non-imaging, 
NFI instruments are the Phoswich Detector System (PDS) which covers $\sim12$ 
to 300 keV (Frontera et al. 1997) and the High-Pressure Gas Scintillation 
Proportional Counter (HP-GSPC) which covers 4 to 120 keV (Manzo et al. 1997). 

A target-of-opportunity observation (TOO) was performed with the NFI four days 
after the WFC detection, on August 26.09-26.61 (i.e., 44.9~ks time span). The net
exposure times are 10.0~ks
for LECS, 23.0~ks for MECS, 12.0~ks for HP-GSPC and 22.2~ks for PDS. 
\bron\ was strongly detected in all instruments and a $\sim$100~s long X-ray burst 
was observed 32.9~ks after the start of the observation. The LECS 
and MECS images show only one bright source, the position as determined from
the MECS image is consistent with that from WFC (Fig.~\ref{figerr} and
Table~\ref{tabpositions}). We applied extraction radii of 8\arcmin\ 
and 4\arcmin\ for photons from LECS and MECS images, encircling at least 
$\sim95$\% of the power of the instrumental point spread function, to obtain 
lightcurves and
spectra. Long archival exposures on empty sky fields were used to define the
background in the same extraction regions. These are standard data sets
made available especially for the purpose of background determination. We note 
that the LECS and MECS backgrounds are not important for such a bright source as
\bron. All spectra were rebinned so as to
sample the spectral full-width at half-maximum resolution by three bins
and to accumulate at least 20 photons per bin. The latter will ensure the
applicability of $\chi^2$ fitting procedures. A systematic error of 1\% was
added to each channel of the rebinned LECS and MECS spectra, to account
for residual systematic uncertainties in the detector calibrations 
(e.g., Guainazzi et al. 1998). The bandpasses were limited to 0.1--4.0 keV
(LECS), 2.2--10.5~keV (MECS), 4.0--25.0 keV (HP-GSPC) and 15--150 keV (PDS)
to avoid photon energies where the spectral calibration of the instruments
is not complete and where no flux was measured above the statistical noise. 
In spectral modeling, an allowance was made to
leave free the relative normalization of the spectra from LECS, PDS and HP-GSPC
to that of the MECS spectrum, to accommodate cross-calibration problems
in this respect. Publicly available instrument response functions were used 
(version September 1997).

\subsection{The persistent emission}
\label{secsppe}

\begin{table}[tb]
\caption[]{Spectral parameters of four acceptable model fits to the persistent 
emission. $\Gamma$ is the photon index. $N_{\rm H}$ is in units of 
10$^{22}$~cm$^{-2}$. The last line of each model specifies the
$\chi^2_{\rm r}$ values for the fit without a bb (black body)
component (first 3 models) or with bb (last model). These values apply
are after re-fitting the remaining parameters (or additional ones for
the last model)}
\begin{tabular}{ll}
\hline
Model              & broken power law + black body\\
$N_{\rm H}$        & $0.86\pm0.04$\\
bb $kT$            & $0.97\pm0.05$ keV \\
$\Gamma_{\rm low}$ & $1.54\pm0.03$ \\
$E_{\rm break}$    & $18.1\pm1.2$ keV\\
$\Gamma_{\rm high}$& $2.13\pm0.04$ \\
$\chi^2_{\rm r}$   & 1.14 (190 dof) \\
$\chi^2_{\rm r}$  without bb  & 1.27 (192 dof)\\
\hline
Model              & high-energy cut off power law \\
                   & ($N(E)(:)E^{-\Gamma}$ for $E<E_{\rm cutoff}$ and\\
		   & $(:)E^{-\Gamma}{\rm exp}(-(E-E_{\rm cutoff})/E_{\rm fold})$\\
		   & for $E>E_{\rm cutoff}$ )\\
                   & + black body\\
$N_{\rm H}$        & $0.82\pm0.04$ \\
bb $kT$            & $0.90\pm0.05$ keV \\
$\Gamma$           & $1.44\pm0.05$ \\
$E_{\rm cutoff}$   & $2.3\pm1.6$ keV\\
$E_{\rm fold}$     & $54.0\pm5.2$ keV\\
$\chi^2_{\rm r}$   & 1.01 (190 dof) \\
$\chi^2_{\rm r}$ without bb & 1.17 (192 dof)\\
\hline
Model              & bremsstrahlung + black body\\
$N_{\rm H}$        & $0.69\pm0.02$  \\
bb $kT$            & $0.84\pm0.02$ keV \\
brems $kT$         & $46.6\pm1.9$ keV\\
$\chi^2_{\rm r}$   & 1.05 (192 dof) \\
$\chi^2_{\rm r}$ without bb  & 3.55 (194 dof) \\
\hline
Model              & Comptonized\\
$N_{\rm H}$        & $0.32\pm0.03$ \\
Wien $kT_{\rm W}$  & $0.57\pm0.01$ keV \\
Plasma $kT_{\rm e}$& $15.5\pm0.6$ keV\\
Plasma optical     & $2.71\pm0.08$ for disk geometry \\
\hspace{2mm}depth $\tau$       & $6.1\pm0.2$ for spherical geometry \\
Comptonization     & 0.9 for disk geometry \\
\hspace{2mm}parameter $y$      & 4.5 for spherical geometry\\
$\chi^2_{\rm r}$   & 1.07 (192 dof) \\
$\chi^2_{\rm r}$ with bb  & 0.97 (190 dof) \\
\hline
\end{tabular}
\label{tabpersistentfit}
\end{table}

\begin{figure}[t]
\psfig{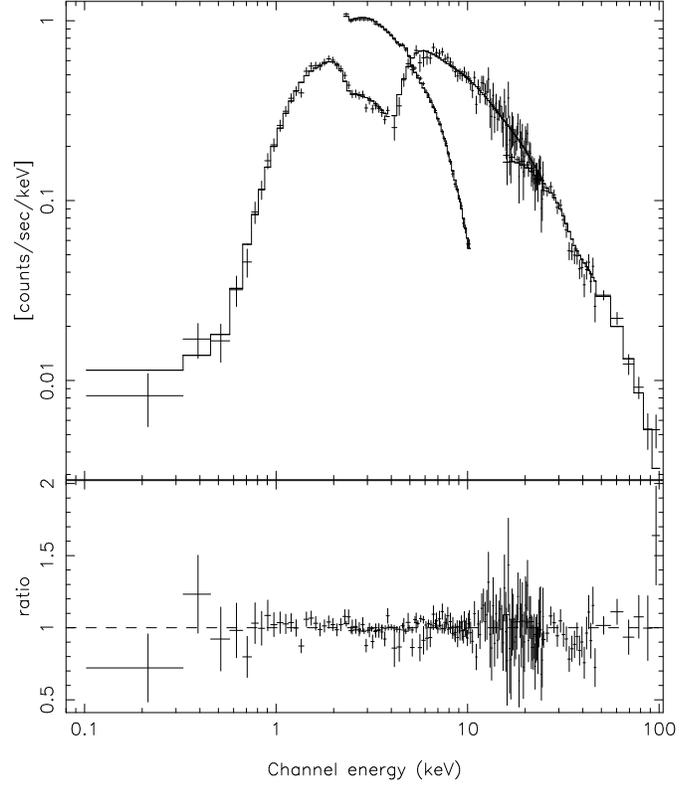}

\caption[]{Upper panel: count rate spectrum (crosses) and Comptonized spectrum
model (histogram) up to 100~keV for average persistent emission. 
Lower panel: residual in units of count rate per channel.
\label{figpersistentspectrum}}
\end{figure}

Fig.~\ref{figpersistentlc} shows the lightcurve in various bandpasses. In the 
two MECS bandpasses a gradual intensity decline is visible by an amount of about
10\%. The statistical quality of the LECS, HP-GSPC and PDS data does not allow a
confirmation of this but these data do not contradict such a decline. 
On top of that, the lowest MECS bandpass shows a $\sim5$\% hump between 27 and 
40~ks into the observation (one might also interpret this as a dip before that).
The X-ray burst occurred at 32.9~ks. Whether there is a physical correspondence
between both is unclear. The statistical quality of the WFC data does not
allow to check this for the three bursts that occurred during that observation.

A broad-band spectrum was accumulated, averaged over the complete observation,
making use of the LECS, MECS, HP-GSPC, and PDS data. As stated above, 
the observation time intervals of the 4 instruments are not exactly overlapping. 
We tested for spectral variations using the MECS data and found no significant
variation. This also applies to the 'hump'. 

A simple absorbed power law function fits the data well between 1 and 40 keV
and the parameter values are consistent with those found for the WFC data
($\chi^2_{\rm r}=1.09$ for 172 dof, the photon index is $\Gamma=1.72\pm0.01$
and $N_{\rm H}=(1.17\pm0.03)\times10^{22}$~cm$^{-2}$).
However, such a model clearly fails when the complete bandpass is considered
($\chi^2_{\rm r}=2.75$ for 196 dof).
In order to explain the spectrum completely, one needs extra emission below 1
keV and a cut off above 40 keV. There are various ways by which this can be
achieved with similar degrees of success. Four successful models are listed in
Table~\ref{tabpersistentfit}. The Comptonized model is according to Titarchuk
(1994)\footnote{We used {\tt XSPEC} as the tool to model spectra, in this
tool the Comptonized model is identified as ``{\tt comptt}''}. It is
parametrized through the temperature $T_{\rm W}$ of the seed photons,
the temperature $T_{\rm e}$ of the scattering electrons, and the optical 
depth of the electron cloud. Two of the four models are successful
in describing 0.5--20~keV spectra of LMXBs in general: black body radiation
plus either thermal bremsstrahlung or a Comptonized spectrum (e.g., White 
et~al. 1989, Christian \& Swank 1997). This supports a LMXB
classification of \bron. 

In the bremsstrahlung model there is the necessity to include a black body
component with a temperature equivalent of $kT=0.84$~keV; the improvement
in $\chi^2_{\rm r}$ is substantial. The other three models also improve 
when including a black body component but the improvement is marginal
according to f-tests. A 1~keV black body feature 
has been detected in a number of other X-ray binary sources as well (e.g., 
Guainazzi et al. 1998, Oosterbroek et al. 1998, Kuulkers et al. 1998, 
Owens et~al. 1997). In some cases it has been modeled by
line emission instead of black body emission.

Independent of the model, the average 0.1-200.0 flux is 
$f_{\rm bol}=1.4\times10^{-9}$~\ecs\ which 
implies, at a distance of 8.5~kpc, a luminosity of $1.2\times10^{37}$~\lum\
(or $3.4\times10^{36}$~\lum\ in 2-10 keV). The 2-10 keV flux is 44\% lower 
than that measured with WFC four days earlier. The column density $N_{\rm H}$ 
is rather insensitive to the model for the continuum except when the 
Comptonized spectrum is considered. It lies between 0.32 and 
0.86$\times10^{22}$~cm$^{-2}$. This range is consistent with 
the value resulting from the interstellar reddening to NGC~6440:
for $E_{\rm B-V}=1.00\pm0.10$ (Ortolani et~al. 1994),
$A_{\rm V}=3.10\pm0.31$ and 
$N_{\rm H}=(1.79\pm0.1)\times10^{21}A_{\rm V}=(5.5\pm0.6)\times10^{21}$~cm$^{-2}$
(according to the conversion of $A_{\rm V}$ to $N_{\rm H}$ by 
Predehl \& Schmitt 1995).

Christian \& Swank (1997) successfully modeled the 0.5 to 20~keV 
spectra of 45 LMXBs by unsaturated Comptonization through a parametrization 
with a cutoff power law. A comparison of the parameter values for \bron\ with 
those found by Christian \& Swank shows that the power law index is normal 
for a burster. The value for $E_{\rm fold}$ is rather high. The values for 
the 8 bursters listed by Christian \& Swank range between 5.3 and 25.0~keV.
However, it should be noted that there is likely to be a selection effect in 
this parameter range because the upper boundary in the data is only 20~keV. 

There are a few bursters which have been measured up to hundreds of keV (e.g., 
KS~1731-260, Barret et al. 1992; 4U~1728-34, Claret et al. 1994; 
Tavani \& Barret 1997). With respect to these bursters, 
the high-energy spectrum of \bron\ is not exceptionally hard. The sample
of bursters with measured broad-band spectra is likely to grow in the near
future; a number of NFI observations of bursters are currently being carried 
out by various investigators. Therefore, in due time, an unambiguous comparative
analysis of all these spectra will be possible. In the mean time,
the NFI study of the burster 1E~1724-308, located in the globular cluster 
Terzan~2 (Guainazzi et al. 1998) provides comparative material (see also, 
e.g., Barret et al. 1999). The NFI spectrum of 1E~1724-308 was successfully modeled 
through the Comptonization model by Titarchuk (1994) plus an additional soft
component in the form of a $\sim1$~keV black body radiator.
The Comptonizing plasma has a temperature of $\sim30$~keV and the seed 
photons have a temperature of 1 to 2~keV. Comparing the temperatures of 
the seed photons and the hot plasma, we note that these are about twice 
as cool in \bron\ than in 1E~1724-308. Also, no additional soft component 
is needed in modeling the spectrum of \bron. We suggest that the differences 
between the two sources are due to a twice as low luminosity in \bron\ which 
could be a reflection of a difference in accretion rate.

\subsection{The burst emission}
\label{secburst}

\begin{figure}[t]
\psfig{figure=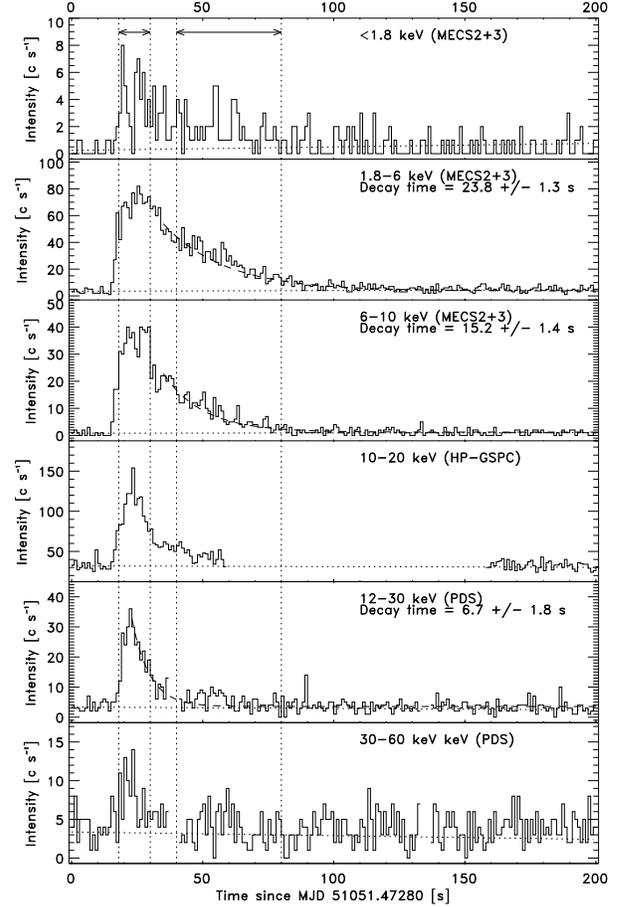,width=\columnwidth,clip=t}

\caption[]{Time profile of the burst detected from \bron\ with MECS, HP-GSPC 
and PDS. The time resolution is 1~s. The curves have not been corrected for
contributions from the background and the persistent emission.
The linear trends as determined from the initial and final 10~s are drawn 
as near-horizontal dashed lines.
The PDS curves were generated from collimator A and collimator B data, the
gaps correspond to collimator slew times between off and on source position
(these last 3~s and occur every 90~s). The HP-GSPC data only include those
for when the collimator was in the on-source position. The vertical lines 
indicate the two time intervals for which spectra were modeled. 
\label{figburstlc}}
\end{figure}

\begin{figure}[t]
\psfig{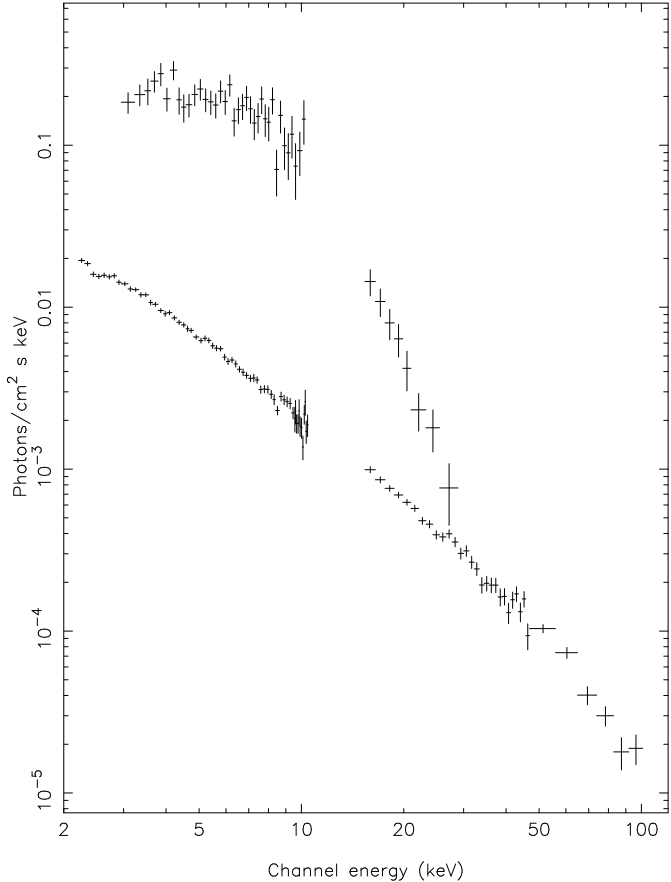}

\caption[]{Photon spectra for the first burst interval (18--30~s, 
upper spectrum)
and for the persistent emission (lower spectrum), for MECS and PDS data only.
\label{figburstspectrum}}
\end{figure}

Fig.~\ref{figburstlc} shows the burst profile in a number of bandpasses
from MECS, HP-GSPC and PDS data. There are no observations of the burst with
the LECS.
The 2 to 10 keV peak intensity is about 0.5 Crab units. This, as well as
the duration, is similar to what was measured for the three bursts
detected with WFC.

The burst was divided in two time intervals (see Fig.~\ref{figburstlc}) and
spectra were generated from MECS (2.2-10.5~keV), HP-GSPC (4.0-15.0~keV)
and PDS data (15.0-30.0~keV) for the first 
interval, and from MECS data (2.2-10.5~keV) for the second interval.
The persistent emission was not subtracted in these spectra. 
We simultaneously modeled the spectra of these two intervals and of the
persistent emission (from all 4 NFI) with the Comptonized model, whose 
parameters were fixed to
the specification in Table~\ref{tabpersistentfit}, plus different 
black body models for the two burst time intervals. The results are given in 
Table~\ref{tabburstfit} and Fig.~\ref{figburstspectrum} illustrates the
spectrum.

\begin{table}[b]
\caption[]{Parameters of single spectral fit to burst and persistent data 
combined; $\chi^2_{\rm r}=1.14$ (398 dof). The first burst interval 
encompasses MECS, HP-GSPC and PDS data, the second only MECS data}
\begin{tabular}{ll}
\hline
bb $kT$ (keV)           & $2.25\pm0.03$ for 18-30 s interval\\
                        & $1.67\pm0.07$ for 40-80 s interval\\
bb radius (km)          & $6.8\pm0.2$   for 18-30 s interval\\
\hspace{0.5cm} for $d_{\rm 8.5~kpc}=1$ & $7.3\pm0.9$   for 40-80 s interval\\
\hline
\end{tabular}
\label{tabburstfit}
\end{table}

The thermal nature of the burst spectrum with few keV temperatures and 
the cooling (Fig.~\ref{figburstlc}, Table~\ref{tabburstfit}) are
typical for a type I X-ray burst (e.g., Lewin et~al.
1995). Such a burst is thought to be due to a thermonuclear
ignition of helium accumulated on the surface of a neutron star. The 
unabsorbed bolometric peak flux of the black body radiation is estimated
at $1.7\times10^{-8}$~\ecs. For a distance of 8.5~kpc this translates into 
a peak luminosity of $1.5\times10^{38}$~\lum. This is near the Eddington 
luminosity for a 1.4~M$_\odot$ neutron star. We do not find evidence for 
photospheric expansion
in this burst as the time profile shows no drops due to adiabatic expansion 
and the black body radius is consistent with a constant value over the two
time intervals used for the spectral analysis. Therefore, we do not expect the 
luminosity to be much higher than the Eddington limit.

Currently, about 50 type I X-ray bursters are known (e.g., Van Paradijs
1995, Heise 1998). About 20 of these have optical counterparts and all 
of those are low-mass stars. 
The remaining bursters do not have optical counterparts because they are
either situated in highly absorbed regions of the sky or are associated
with transient X-ray sources whose position was not measured with sufficient
accuracy.
It is very likely that \bron\ is a low-mass X-ray binary as well, 
especially since it appears associated with a globular cluster.

In the PDS, the photon count rate of the black body radiation is estimated
to be negligible between 30 and 60 keV. For a $kT=2.25$~keV black body model
it is $4\times10^{-3}$ times that in 12 to 30 keV. Nevertheless, the 30-60 keV 
PDS bandpass shows an increase during the first burst interval 
(Fig.~\ref{figburstlc}). In this interval, the average of the increase in 
30-60~keV photon count rate is $4.0\pm0.5$~c~s$^{-1}$ which
compares to a level outside the burst interval of $1.01\pm0.03$~c~s$^{-1}$.
The ratio of the average 30-60~keV through 12-30~keV photon count rate 
increase during the first burst interval is $0.21\pm0.03$. 
We are unable to determine the $>30$~keV spectrum during this interval due 
to insufficient statistics. If one assumes that the spectrum is the same as
outside the burst, it implies that the level of Comptonization during the
first burst interval increased by a factor of $\sim4$.
The close match between the $<30$ and $>30$~keV time profiles
(Fig.~\ref{figburstlc}) indicates that the presumed plasma 
is within $\sim$10$^{11}$~cm of the neutron star surface which is the location
of the black body radiation.

We note that the PDS data are crucial to observe the Comptonization
of the burst emission. The MECS data by itself can satisfactorily be modeled
by black body radiation only with $N_{\rm H}$ fixed at the value found from
the persistent spectrum. For the first burst interval, $\chi^2_{\rm r}=0.94$ 
for 35 dof and the temperature is slightly higher (2.37~keV). 

No other bursts were observed during the NFI interval; this excludes
that the time interval between bursts during the NFI observation is
the same as that observed between the bursts during the WFC observation, 
in which case two more bursts should have been detected.

\section{Rapid variability measurements with {\em RossiXTE\/}-PCA}
\label{secvar}

\begin{figure}[t]
\psfig{figure=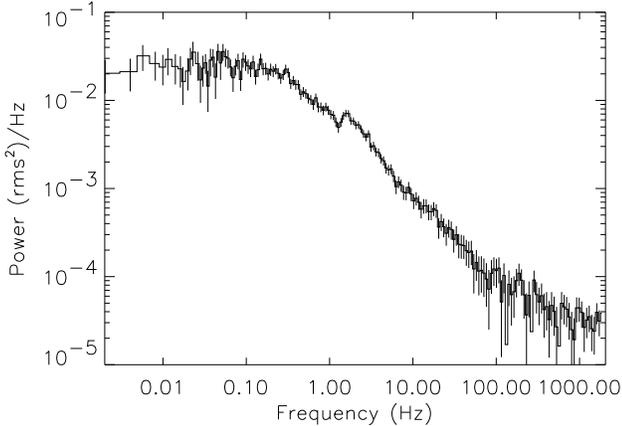,width=\columnwidth,clip=t}

\caption[]{Fourier power spectrum of the intensity time series as measured with
{\em RossiXTE\/}-PCA between 2 and 20~keV
\label{figpower}}
\end{figure}

A TOO was performed with {\em RossiXTE\/} on August 26.28--26.57, 4 days after the WFC 
detection. This TOO was nearly simultaneous to the one with NFI but the actual 
overlap with live NFI/MECS time is less than the exposure time, due to the 
different orbits of the two spacecrafts. The X-ray burst
detected with the {\em BeppoSAX\/} NFI was missed by {\em RossiXTE\/} and no other bursts were
detected with {\em RossiXTE}. The power spectrum of the 2 to 20~keV {\em RossiXTE\/}
Proportional Counter Array (Zhang et al. 1993) time series shows no significant
narrow features. The 90\% confidence upper limit to any periodic signal power in
the 100 - 900 Hz range is 0.37\% (rms). The broad-band power spectrum computed from
9.7 ksec of data in the 2 to 20~keV range is shown in Fig.~\ref{figpower}.
It is typical of LMXBs of lower luminosity (i.e., with a luminosity at least one
order of magnitude below the Eddington limit), being dominated
by a power law above 0.2 Hz and flattening below. We fitted this spectrum
with two components, a broken power-law and a "bump" between 1 - 2 Hz which
we have modeled with a burst-like profile (linear rise, exponential decay). 
The derived power law indices, break
frequency, location of the few Hz "bump", and the rms amplitudes of the components
(see Table~\ref{tabpow})
are all consistent with broad-band power spectral measurements of other
LMXB of lower luminosity (e.g., Wijnands \& Van der Klis 1998a).

\begin{table}
\caption[]{Model parameters of broken power law fit to PCA Fourier power 
spectrum\label{tabpow}}
\begin{tabular}{ll}
Parameter & Value \\
\hline
Low-frequency index    & $0.058 \pm 0.032$ \\
High-frequency index   & $0.940 \pm 0.015$ \\
Break frequency        & $0.29 \pm 0.02$ Hz \\
rms (0.001 to 2048 Hz) & $4.8 \pm 0.3$\%  \\
Bump start frequency   & $1.31 \pm 0.06$ Hz \\
Bump peak frequency    & $1.54 \pm 0.07$ Hz \\
Bump decay frequency   & $1.4 \pm 0.1$ Hz \\
rms (Bump start to 2048 Hz) & $6.8 \pm 0.8$\% \\
\hline
\end{tabular}
\end{table}

\section{Discussion} 
\label{secdis}

The position, flux, spectrum and rapid variability are fully consistent with
\bron\ being the immediate neighborhood of a neutron star as part of a 
LMXB inside the Galactic globular cluster NGC~6440. The 
X-ray luminosity outside the burst is $\sim10^{37}$~erg~s$^{-1}$
which classifies the source as a LMXB of lower luminosity. This is consistent
with the source being a type I burster and having a hard spectrum. It shows
characteristics of an Atoll-type source (Hasinger \& Van der Klis 1989, Van der 
Klis 1989).

The persistent spectrum can be well described by a number of models. In
general, spectra of LMXBs of lower luminosity can be explained by
a black body plus either thermal bremsstrahlung or a Comptonized spectrum.
For \bron, both models fit the data equally well. Still, the
most promising appears to be the Comptonized model whereby the spectrum
is explained by soft photons that are Compton
up scattered by a $\sim15$~keV hot plasma. The reason that this model is
preferred is because during a thermonuclear explosion on the neutron star
the hard emission brightens simultaneously and dies out in
the same fashion. The natural explanation of this is that photons created
during the explosion are inverse-Compton scattered immediately by the hot plasma
which is within $\sim10^{11}$~cm of the neutron star. In case of thermal
bremsstrahlung, the brightness would be linked to the mass accretion rate. In a
thermonuclear flash it is not expected that the accretion rate increases and the 
thermal bremsstrahlung brightens.

Where is the source of the soft non-burst photons located? In order to obtain an
expression for the equivalent spherical radius $R_{\rm W}$ of the emission area
of the Wien spectrum, one needs to equalize the bolometric luminosity of the
soft photons, $L_{\rm pers}$, to that of a black body with temperature 
$kT_{\rm W}$ and radius $R_{\rm W}$. $L_{\rm pers}$ can be obtained from 
integrating the observed
spectrum ($f_{\rm bol}$) and the distance $d$ to the source. One will have 
to correct for energy gained by the photons in the inverse Compton scattering. 
The relative gain is given by the Comptonization parameter 
$y=4kT_{\rm e}\tau^2/m_{\rm e}c^2$. $R_{\rm W}$ is then given by
$R_{\rm W}=3\times10^4 d \sqrt{\frac{f_{\rm bol}}{1+y}}/(kT_{\rm W})^2$~km
where $d$ is in kpc, $f_{\rm bol}$ in \ecs, and $kT_{\rm W}$ in keV. If we 
apply the values of $f_{\rm bol}=1.4\times10^{-9}$~\ecs, $y=4.5$ (for spherical 
geometry) and $kT_{\rm W}=0.57$~keV, 
then $R_{\rm W}\sim13d_{\rm 8.5~kpc}$~km.
This implies that the soft photons are generated at the boundary layer between 
the accretion disk and the neutron star surface. The hot plasma is probably located
to the outside of this layer because otherwise the spectrum would have been
thermalized. The bolometric peak flux of the burst is
$2.1\times10^{-8}$~\ecs, $\sim$20\% of this constitutes scattered photons.

The behavior of the Comptonized spectrum during the burst has been observed 
in one other LMXB before the launch of {\em BeppoSAX}: X1608--52 (Nakamura et al., 
1989) with an instrument with a bandpass of 2 to 20 keV. The NFI on {\em BeppoSAX\/} 
in principle provide better means to test the Comptonization model on bursts 
because they combine to such a broad bandpass that the $kT$ values for the 
Wien and plasma temperatures are within the bandpass. However, it is crucial 
that PDS data are available during the burst.
Apart from the analysis presented here, we are aware of an NFI analysis of 
an X-ray burst in one other LMXB reaching similar conclusions: GS~1826-24
(Del Sordo et al. 1998).

On the basis of a short ROSAT HRI observation Johnston et al. (1995) report
one dim ($L(0.5-2.5\,{\rm keV})\simeq 1.2\times10^{33}$\,\lum) X-ray
source near the core of NGC6440; a later longer observation indicates the
presence of two sources (Verbunt 1998). At the moment
it is not clear whether the transient corresponds to either of them;
a ROSAT observation obtained on September 8, 1998, may shed light on this
(Verbunt et al., in preparation).
Considering the fact that no globular cluster contains two bright sources,
we think it rather likely that \bron\ is the same source as
MX\,1746-20 though this is not completely supported by the characteristics. 
The 2-10\,keV spectrum of MX\,1746-20 fits a power law with
$\Gamma=2.1_{-0.5}^{+0.8}$ and 
$N_{\rm H}=(5.3_{-2.0}^{+3.0})\times10^{22}$~cm$^{-2}$
(Markert et al.\ 1975). The photon index is compatible with the NFI spectrum of 
\bron\ in that energy range but $N_{\rm H}$ is inconsistent.
The peak flux of MX\,1746-20 was about 150 mCrab, about five times
brighter than the WFC detection of Aug 22, 1998.
Such a variation of peak fluxes is not uncommon among recurrent transients
(e.g., Chen et~al. 1997, Levine 1998).

\bron\ declined $44\pm5$\% in 3.6 days. This is equivalent 
to an e-folding decay time of $6\pm1$~d. If the source
declined exponentially, the intensity would
have been 6~mCrab during the WFC observation on Sept. 1, equivalent with a 4$\sigma$
signal. This is not inconsistent with the non-detection, as a 4$\sigma$ signal
after randomization might end up below a 3$\sigma$ detection level. The $10\pm2$\%
drop in the 2-10 keV intensity during the NFI observation is equivalent
to an e-folding decay time of $5\pm1$~d, in good agreement
with the decay time above. For a decay time of 6~d and an activity onset 
4 days before the WFC detection, as suggested by the ASM measurements,
the peak flux would roughly be 57~mCrab. The 7-day average flux measurement
with ASM is consistent with that.

Faint transient LMXBs that last only a few weeks are being discovered more and 
more,
thanks to instruments on {\em BeppoSAX\/} and {\em RossiXTE\/} (Heise 1998, Remillard 1998). 
These short-period transients appear characterized by a peak luminosity well 
below the Eddington luminosity (by one to two orders of magnitude).
\bron\ belongs to this class of transients.
The most appealing example of such a transient is SAX~J1808.4--3658 (In 't Zand
et al. 1998c) which revealed a pulsar signal and thus was determined to be in a 
tight binary system with a projected semi-major axis of only $1.9\times10^{9}$~cm 
(Chakrabarti \& Morgan 1998, Wijnands \& Van der Klis 1998b). One could 
speculate that the small outbursts of these transients are the 
consequence of a small accretion disk as fits in a short-period binary 
system. Confirmation of such speculation requires direct measurements
of the orbital parameters of the binary.

Currently, 
bright (i.e., with an X-ray luminosity in excess of 10$^{36}$~\lum) X-ray 
sources are known in 12 Galactic globular clusters: in NGC~1851, 6440, 6441, 6624, 
6652, 6712, 7078, Terzan 1, 2, 5, 6, and Liller 1. The sources in five of these 
are transient (e.g., Verbunt et al. 1995). The detection of X-ray bursts from 
\bron\ leaves only one bright globular cluster source from which no X-ray bursts 
have been detected: Terzan 6 which contains one of the transients. The fraction 
of the number of type I X-ray bursters in low-mass X-ray binaries
in the Galactic disk is 30 to 40\% (Van Paradijs 1995, Heise 1998), that in Galactic
globular clusters is 92\%. This difference was noted before and was partly suspected
to be due to selection effects (infrequent and far-away bursters in the disk
may easily go unnoticed). However, with the advent of WFC and its extensive 
program on the Galactic bulge, these selection effects have decreased
considerably.

It has been suggested (e.g., In~'t~Zand et al. 1998b) that another difference 
between the LMXB populations in the disk and the globular clusters may be the 
fraction of transient systems which contain black hole candidates. For the disk 
it is at least 20\%  (Tanaka \& Shibazaki 1995) and possibly as high as 70\% 
(Chen et al. 1997). For a population of five LMXB transients in the 
clusters this percentage now is smaller than 20\%. Assuming that 70\% is the true 
value for both populations, the chance probability of finding no black hole candidate
among the five cluster transients is 3\%. Although intriguing, this number is 
not conclusive. Also, it is subject to uncertainty. For instance, as Deutsch et~al.
(1998) pointed out for NGC~6652, a classification as 
transient is only indisputable if a source varies by several orders of magnitudes
(we note that similar classification problems exist for transients in the Galactic
disk, e.g., Chen et al. 1997).
The count of 5 transient LMXBs is based on a classification that flags a source 
as transient when it is usually below 10$^{36}$~\lum\ and occasionally 
brightens to above that threshold. This definition may include sources that
vary less than orders of magnitude but happen to have an average luminosity
close to the threshold. The source in NGC~6652 falls into that category. 
\bron\ in NGC~6440 does not, it is four orders of magnitude above the quiescent
level measured with ROSAT (Johnston et al. 1995). The only other
indisputable transients are Terzan 6 and Liller 1. The chance 
probability of finding no black hole candidates among 3 transients is 12\%.

\begin{acknowledgements}
We thank Jacco Vink and Tim Oosterbroek for help in the use of the
analysis software, the {\em BeppoSAX\/} team at Nuova Telespazio (Rome) for
planning the observations and processing the data, and Gerrit Wiersma,
Jaap Schuurmans and Anton Klumper for help in the analysis of the WFC data. 
{\em BeppoSAX\/} is a joint Italian and Dutch program.

\end{acknowledgements}

\end{document}